\documentclass[submission,copyright,creativecommons,noncommercial]{eptcs}
\providecommand{\event}{MARS 2024} % Name of the event you are submitting to

\usepackage{iftex}

\ifpdf
  \usepackage{underscore}         % Only needed if you use pdflatex.
  \usepackage[T1]{fontenc}        % Recommended with pdflatex
\else
  \usepackage{breakurl}           % Not needed if you use pdflatex only.
\fi
\usepackage{amsmath,amssymb}
\usepackage{eucal}
\usepackage{subcaption,graphicx}

\usepackage{tabularray}
\usepackage{array}

\usepackage{listings}
\usepackage{tikz}
\usetikzlibrary{arrows}

\makeatletter
\newcommand\@erelb@r[1]{%
  \mathrel{\tikz[baseline=-.5ex]\draw[#1] (0,0)--(0.3,0);}
}
\newcommand{\erelbar}[1]{\@erelbar#1}
\def\@erelbar#1#2{%
  \ifcase\numexpr#1*4+#2\relax
    \@erelb@r{-}\or     % 00
    \@erelb@r{->}\or    % 01
    \@erelb@r{-|}\or    % 02
    \@erelb@r{->|}\or   % 03
    \@erelb@r{<-}\or    % 10
    \@erelb@r{<->}\or   % 11
    \@erelb@r{<-|}\or   % 12
    \@erelb@r{<->}\or   % 13
    \@erelb@r{|-}\or    % 20
    \@erelb@r{|->}\or   % 21
    \@erelb@r{|-|}\or   % 22
    \@erelb@r{|<->|}\or % 23
    \@erelb@r{|<-}\or   % 30
    \@erelb@r{|<->}\or  % 31
    \@erelb@r{|<-|}\or  % 32
    \@erelb@r{|<->|}    % 33
  \else
    \@wrong
  \fi
}
\makeatother
\newcommand{\xrightarrowdash}[1]{\overset{#1}{\erelbar{21}}}

\makeatletter
\newcommand{\oset}[3][0ex]{%
  \mathrel{\mathop{#3}\limits^{
    \vbox to#1{\kern-2\ex@
    \hbox{$\scriptstyle#2$}\vss}}}}
\DeclareRobustCommand{\sqcdot}{\mathbin{\mathpalette\morphic@sqcdot\relax}}
\newcommand{\morphic@sqcdot}[2]{%
  \sbox\z@{$\m@th#1\centerdot$}%
  \ht\z@=.33333\ht\z@
  \vcenter{\box\z@}%
}
\makeatother
\newcommand{\dat}[1]{\oset[-.5ex]{#1}{-\!\!\sqcdot\,}}
\newcommand{\ok}{\mathbf{ OK}}

\title{Formally Modelling the Rijkswaterstaat Tunnel Control Systems in a Constrained Industrial Environment}
\author{Kevin H.J.~Jilissen
\institute{Rijkswaterstaat\\ Utrecht, the Netherlands}
\institute{Eindhoven University of Technology\\
Eindhoven, the Netherlands}
\email{kevin.jilissen@rws.nl}
\and
Peter Dieleman
\institute{Rijkswaterstaat\\ Utrecht, the Netherlands}
\email{peter.dieleman@rws.nl}
\and
Jan Friso Groote
\institute{Eindhoven University of Technology\\
Eindhoven, the Netherlands}
\email{j.f.groote@tue.nl}
}
\def\titlerunning{Formally Modelling the RWS Tunnel Control Systems in a Constrained Industrial Environment}
\def\authorrunning{K.H.J.~Jilissen, P.~Dieleman \& J.F.~Groote}
\begin{document}
\maketitle

\begin{abstract}
  Rijkswaterstaat, the National Dutch body responsible for infrastructure, recognised the importance of formal 
  modelling and set up a program to model the control of road tunnels.
  This is done to improve the standardisation of tunnel control and  
  make communication with suppliers smoother. A subset of SysML is used to formulate the models, which are substantial.
  In an earlier paper we have shown that these models can be used to prove behavioural properties by manually translating
  the models to mCRL2. In this paper we report on an automatic translation to mCRL2.
  As the results of the translation became unwieldy, we also investigated modelling tunnel control 
  in the specification language Dezyne which
  has built-in verification capabilities and compared the results.
\end{abstract}

\section{Introduction}
Over the last few years, Rijkswaterstaat (RWS, the Dutch body responsible for road and water infrastructure in the Netherlands) 
has created SysML models of all system parts of the tunnel control systems describing the functionality these systems should perform based on a functional decomposition.
They use SysML as within RWS there is a preference to use modelling tools with commercial support and industry acceptance. 
These models have been created as generic blueprints for the construction of several road tunnels.
The behaviour of the systems is modelled using a functional decomposition in nested Activity Diagrams.
There is no formal semantics to which the models adhere and essential parts are denoted in `structured natural language'.

Formal methods could be applied to reduce the chance of system failures and the chance that there are design flaws in the systems.
RWS has shown interest in this approach, and in \cite{jilissen2023formal} it is already shown that a structured but largely manual 
translation of these SysML models to the formal mCRL2 \cite{bunte2019mcrl2} specification language is possible and very beneficial.
The goal of this translation is to improve the quality of the models by formal verification of both safety and liveness properties of 
both the overall system as well as its individual components.

However, a manual translation to formal models is not deemed as the desirable solution by RWS as the 
set of models is substantial and they are regularly revised and changed. Manual translation is time consuming 
and relatively error prone. This may lead to the situation that changes in the SysML models quickly become out of sync
with the verifiable mCRL2 models leading to a reduced benefit of verification. 

For this reason, we investigate if the existing SysML specifications can automatically be translated to mCRL2.
The first major obstacle is that SysML has no formal semantics --- but see for instance \cite{lima2013formal} 
on how this can be remedied  --- and in particular the
`structured natural language' in the SysML models
cannot be formally interpreted and translated. Therefore, we first systematically translate this to 
SysML activity diagrams. 
Subsequently, based on the XML interchange format, we constructed a translator in Spoofax \cite{kats2010spoofax}, which
is a language translation workbench with built-in support for variable declarations with local scope. 

The communication scheme employed in the SysML models of tunnels is that all components simultaneously read their 
input and deliver their output.
In \cite{jilissen2023formal} it was already observed that this leads to mCRL2 models with relatively few states but with a huge number
of outgoing transitions in each state, sometimes more than $10^8$. The automatic translation showed that this problem was
exacerbated with the larger models making the verification of properties cumbersome. 

Therefore, we also investigated modelling tunnels in the Dezyne specification language \cite{beusekom2021dezyne}, 
which is commercially available and 
has very effective built-in verification. 
Following guidelines in \cite{groote2015specification}, we provide both push and pull models for 
certain tunnel
control components, and compared the behaviour with those generated from the SysML description, which was still possible
despite the very different styles of modelling. 

This paper presents the journey to develop support to improve the quality of the semi-formally specified tunnel control systems, 
which is a project largely running within Rijkswaterstaat. 
The results are in no way as clear cut as we hoped, and this work led to some conclusions, quite different from what we originally
anticipated. 
We believe that these conclusions can be generalised to other behavioural specifications in graphical languages without a proper formal semantics.
Besides this project, the enhancement of safety and reliability using formal methods is also being 
investigated by employing Synthesis Based Engineering (SBE) \cite{moormann2022light}. In this approach formal safety properties
are used to automatically generate the control models, which is quite different from our approach to verify liveness and safety
properties on explicitly specified control systems.

\section{Existing SysML model structure}
The complete tunnel installation has been generically modelled in Enterprise Architect \cite{sparx0000enterprise} using SysML version 1.5 \cite{omg2015systems} and documented respecting J-STD-16 \cite{J-STD-016-1995}. The genericity is introduced by a parameterised description that can be instantiated for any specific tunnel configuration.
Examples of such parameters include the number of traffic tubes, the number of lanes, and the number and configuration of ventilators.

The permitted model elements and relations are specified using SysML in the meta-model of the model.
This meta-model prescribes that all functionalities defined in the Landelijke Tunnel Standaard (translated: Dutch National Tunnel Standard) \cite{rijkswaterstaat2021landelijke} must be modelled as behaviour using Activity Diagrams (ADs).
In these ADs, the repeated token flow simulates the computation cycles of the Programmable Logic Controllers (PLCs) deployed in tunnel installations.

\begin{figure}[!ht]
  \begin{subfigure}{.59\textwidth}
    \centering
    \includegraphics[width=.98\linewidth]{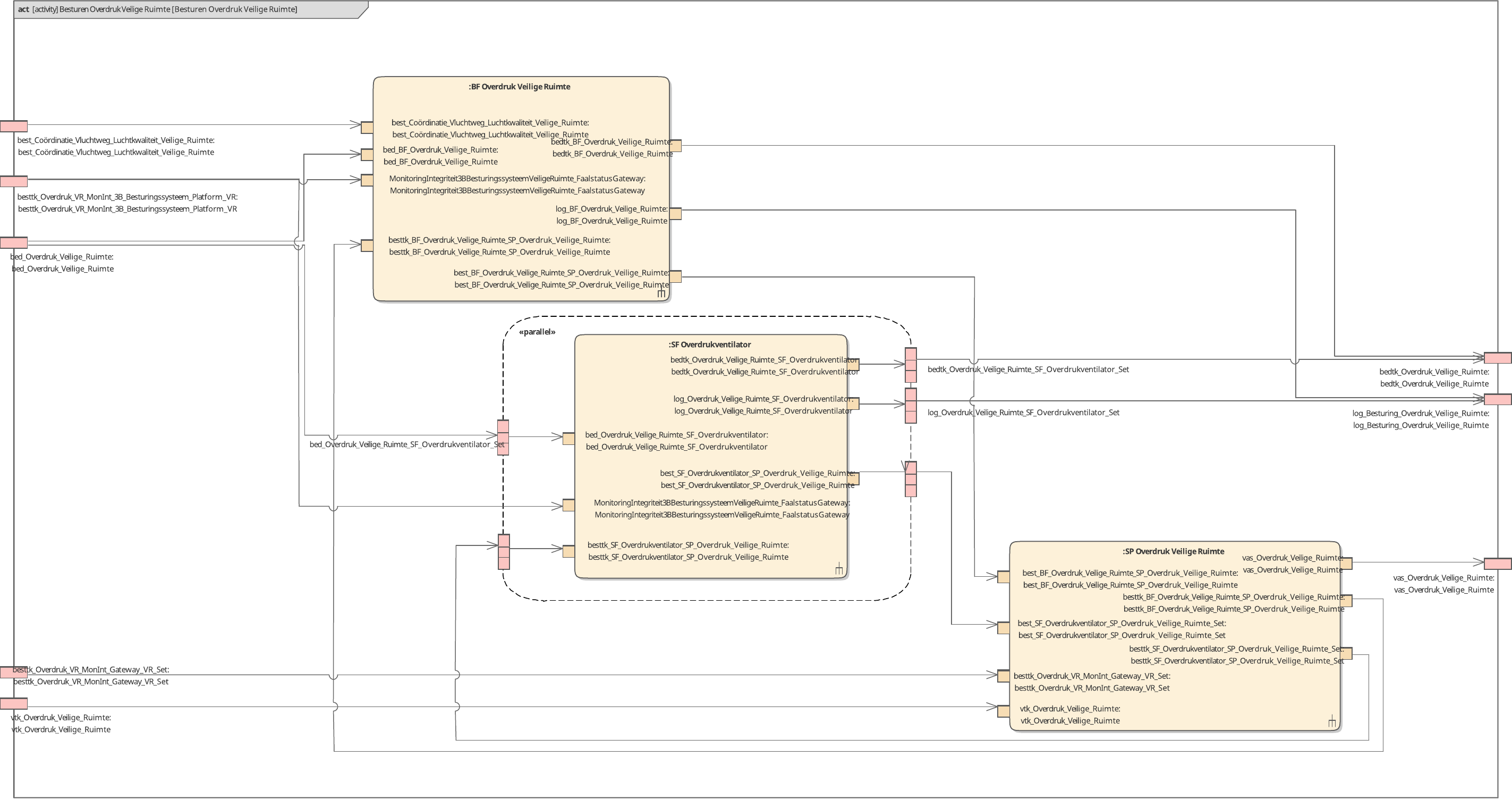}
    \caption{The leaf AD of the overpressure sub-system.}
    \label{fig:existing-ad}
  \end{subfigure}
  \begin{subfigure}{.39\textwidth}
    \centering
    \includegraphics[width=.98\linewidth]{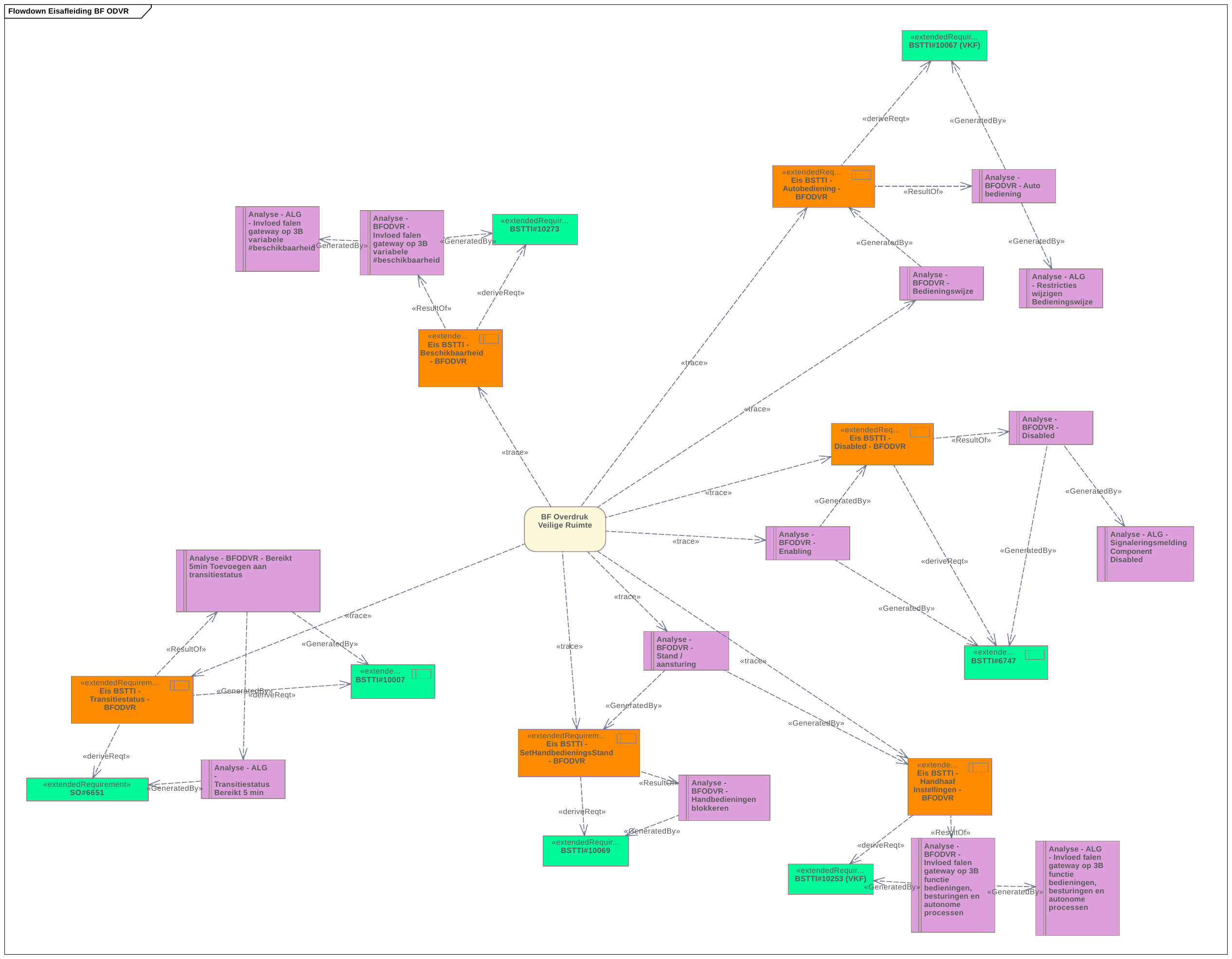}
    \caption{Additional descriptions for Figure \ref{fig:existing-ad}.}
    \label{fig:existing-link}
  \end{subfigure}
  \\
  \begin{subfigure}{0.59\textwidth}
    \centering
    \begin{lstlisting}
  EnableOverpressure()
  Condition: *
  Actions: #enabled := yes
    \end{lstlisting}
    \caption{An elementary task description translated from Dutch.}
    \label{fig:existing-req}
  \end{subfigure}
  \begin{subfigure}{0.39\textwidth}
    \centering
    \begin{lstlisting}
  DisableOverpressure()
  Condition: *
  Actions: #enabled := no
    \end{lstlisting}
    \caption{Another elementary task description.}
    \label{fig:existing-req2}
  \end{subfigure}
  \caption{An example of a leaf AD taken from the SysML description of tunnel control systems.}
  \label{fig:existing}
\end{figure}

The behaviour of the tunnel control system is modelled as a functional decomposition of Activity Diagrams.
The root AD contains one component encapsulating all behaviour of the system, with the environmental 
readings as input and the actuator control as output.
Every behavioural component in the system contains subcomponents 
until every technical or system task is described by exactly one leaf subcomponent.
An example of such leaf AD containing elementary tasks, or activities, 
regarding controlling the overpressure of safe spaces, in Dutch `overdruk veilige ruimte', 
is depicted in Figure \ref{fig:existing-ad}. The elementary tasks are subject to various further
descriptions in SysML, as depicted in Figure \ref{fig:existing-link}. 
Important sub-descriptions are how actions, consisting of value assignments,
must be carried out subject to certain conditions when the system receives messages, see Figure \ref{fig:existing-req}.
These descriptions consist of a curious mixture of Dutch and programming notation, to 
which we refer in this document as `structured natural language', which has no formal syntax and semantics. 
Note that the diagrams are often
small and cluttered, and require the zoom feature of pdf to be readable.

Basic functionalities in the model are grouped together in so-called vertical slices. 
In total, the engineers identified 35 vertical slices in the systems.
These slices together contain 52 standalone sub-systems responsible for some facility.
The sub-systems are decomposed in a Base Functionality (BF) responsible for the overall management of that sub-system.
They can contain zero or more Sub-Functionalities (SF) responsible for a single, possibly instantiated, entity, such as a single ventilation group or a single ventilator.
Finally, the sub-systems can contain Drivers (SP, abbreviation of `Stuurprogramma' in Dutch) which translate the generic control commands and status information to and from the vendor-specific implementations.

For this paper, the vertical slice and equally named sub-system for the overpressure of the safe space 
is used as the system we apply our techniques on.
It consists of a Base Functionality and one Sub-Functionality which can be instantiated for (possibly redundant) overpressure ventilators.
This sub-system was chosen as it is one of the investigated systems in \cite{jilissen2023formal}.
Additional details on translations and models are provided in
the appendices. In Appendix \ref{app:artefacts} a non traceable link is given to a zip file containing
all artefacts belonging to this paper.

\section{Model adaptions for formal analysis}
\begin{figure}[!ht]
  \centering
  \includegraphics[width=.95\linewidth]{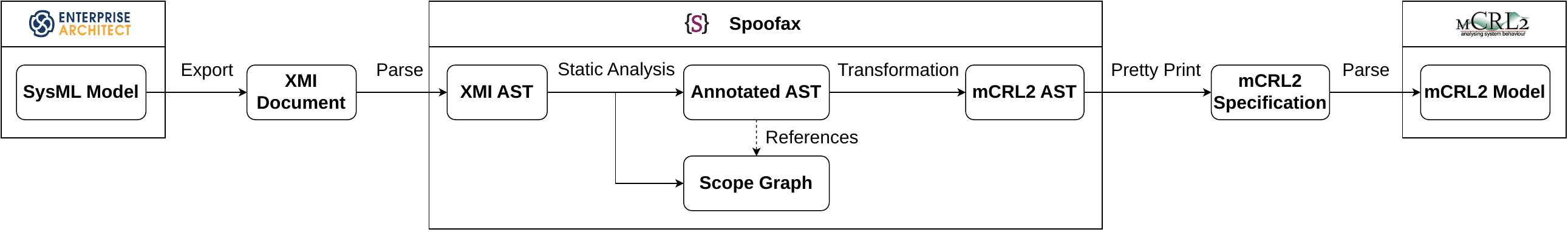}
  \caption{An overview of the workflow.}
  \label{fig:workflow}
\end{figure}

In this section we sketch how the SysML models are translated to mCRL2.
We only allow a restricted use of SysML activity diagrams, and this is enforced by the type checking in the translation.
The transformation is implemented using the Spoofax Language Workbench \cite{kats2010spoofax}.
Syntax definitions for parsing XMI are described in Spoofax using the SDF3 meta-language \cite{souza2020multi}.
Static analysis rules on the parsed Abstract Syntax Tree (AST) are formulated in Statix \cite{antwerpen2018scopes}, the meta-language for the specification of static semantics included in Spoofax 3.
The analysed input AST is transformed using Stratego \cite{kalleberg2007spoofax} 
to an AST of the resulting mCRL2 specification and exported.
Figure \ref{fig:workflow} gives an overview of this workflow.

\subsection{Formalising `structured natural language'}
The first step towards the translation is to replace the structured natural
language descriptions by a notation that can be formally interpreted and 
translated. We formalise the 
structured natural language descriptions using activity diagrams annotated with simple conditions and assignments in SysML, as this deviates the least from the framework Rijkswaterstaat is
using.

\begin{figure}[!ht]
  \begin{subfigure}{.39\textwidth}
    \centering
    \includegraphics[width=.98\linewidth]{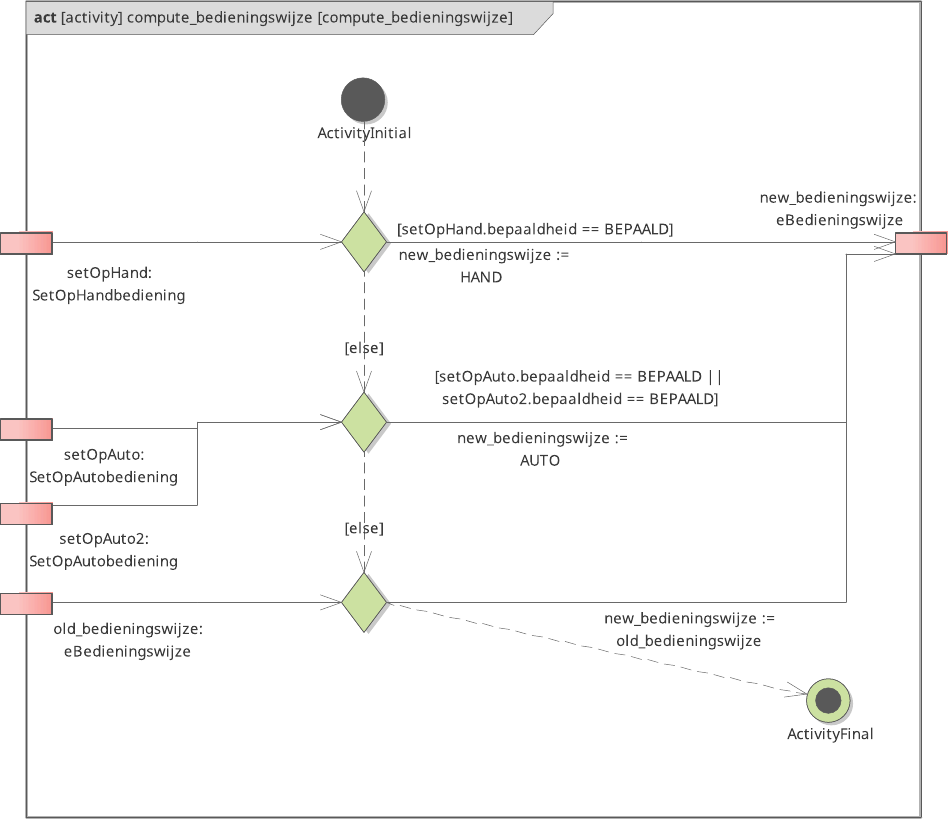}
    \caption{Example AD in the leaf layer.}
    \label{fig:leaf-ad}
  \end{subfigure}
  \begin{subfigure}{0.59\textwidth}
    \centering
    \includegraphics[width=\linewidth]{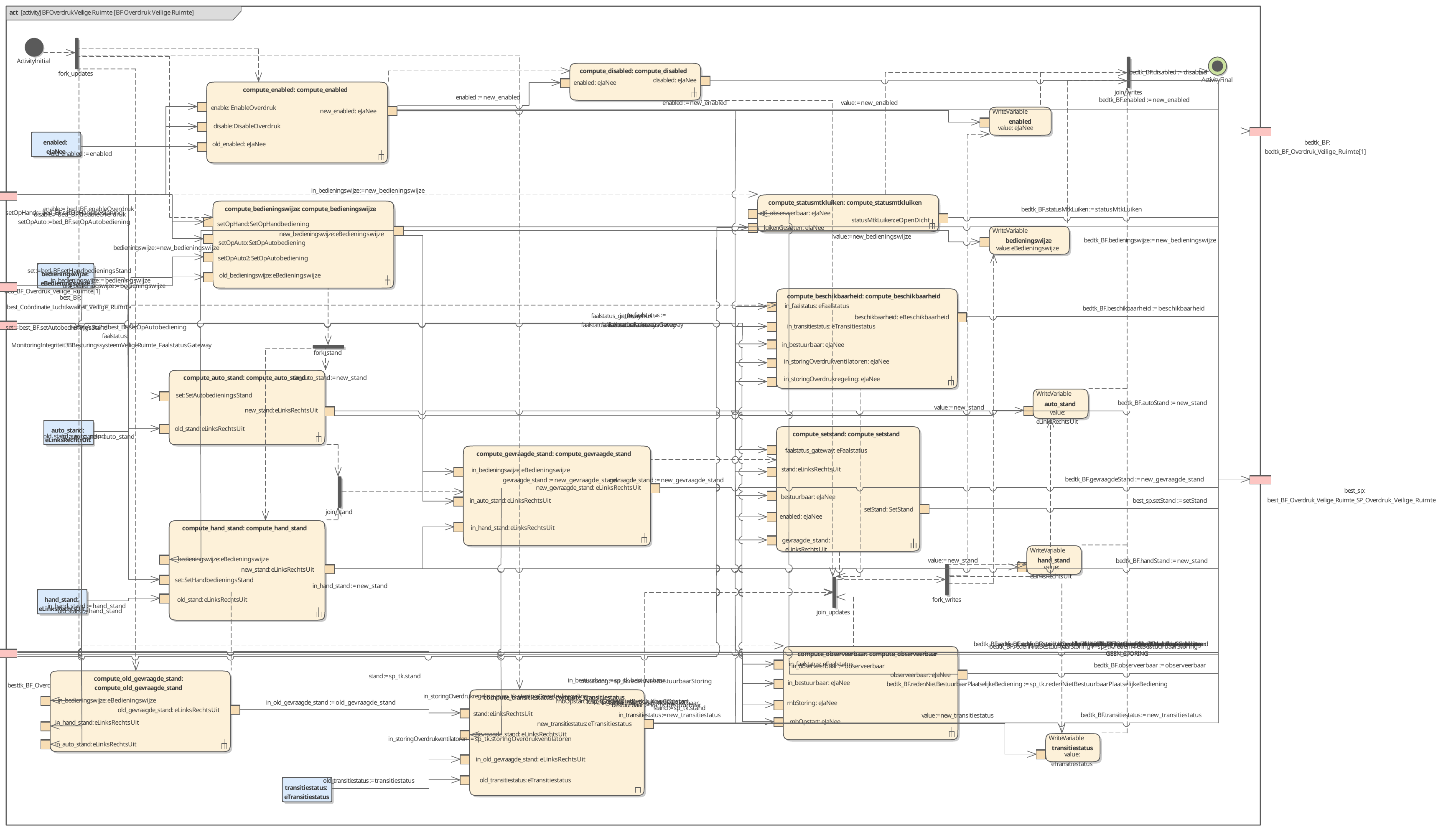}
    \caption{Example AD in the glue layer.}
    \label{fig:glue-ad}
  \end{subfigure}
  \caption{Example Activity Diagrams in the newly introduced layers.}
  \label{fig:new-ads}
\end{figure}

To create this description, two additional decomposition layers must be added.
These manually created layers embed the structured natural language with a fixed semantics within the SysML model, such that these can be constructed and maintained by the SysML engineers.
A new bottom leaf layer consists of ADs, in which each AD formally describes how several task descriptions such as in Figures \ref{fig:existing-req} and \ref{fig:existing-req2} combined compute the values of the assigned
variables. An example of an AD in the leaf layer is given in Figure \ref{fig:leaf-ad}.

In between this newly created leaf layer 
and the ADs of the existing SysML specification, 
such as Figure \ref{fig:existing-ad}, a glue layer is added.
For each activity in the existing AD, an AD in the glue layer is added which formally describes how the variables in the textual descriptions are bound to input, output, and state variables.
The description in Figure \ref{fig:existing-ad} requires three ADs in the
glue layer of which  
one is depicted in Figure \ref{fig:glue-ad}.

\subsection{Assignment-based language in SysML}
\label{sec:assignment}
SysML allows modellers to define their own language to be used as names and guards on flows in diagrams.
In the ADs in Figure \ref{fig:leaf-ad}, a simple conditional assignment-based language is used which is just expressive enough to capture the natural language encountered in the specifications.
The design for this language is guided by minimalism,
which is trivial to transform to mCRL2.

The goal of the language is to assign values to all outgoing activity parameters and to the pins of all actions that write a new value to a state variable.
For this purpose, flow names become assignments of the shape $e_1 := e_2$ 
and flow guards are boolean expressions with connectives 
$\neg$, $\vee$ and $\wedge$, and basic propositions of the shape 
$e_1 = e_2$ and $e_1 \not= e_2$.
Within a sensible name resolution scope of the flow, $e_1$ and $e_2$ must refer to named 
elements and values of equivalent types in the SysML model.

Defining a sensible name resolution scheme and determining what equally typed values are is the most 
complex task of the transformation from XMI to mCRL2.
For this we use the tool Statix, in which type checking is reduced to a constraint solving problem.
If a solution is derived for the constraints on the root in the AST, the provided AST is well-typed.
A scope graph is constructed in Statix while solving this constraint problem \cite{antwerpen2018scopes,neron2015theory}.

The scope graph framework consists of a graph, which represents the naming structure of the AST, and a resolution calculus, which describes how to resolve references to declarations within a scope graph.
A scope graph $\mathcal{G}$ connects \textit{scopes} $s \in \mathcal{S}$, containing \textit{data terms} $d \in \mathcal{D}$ bound by \textit{relations} $r \in \mathcal{R}$, using directed \textit{edges} labelled with \textit{labels} $l \in \mathcal{L}$.
A scoped datum $s \dat{r} d$ associates a data term $d$ with scope $s$ under relation $r$.
Variable declarations in scope $s$ are represented by $s \dat{\colon} (n, T)$, shortened by $n \colon T$ for name $n$ with type $T$.
Resolution between scopes is governed by path queries.
The query can be read as a regular expression with the edge labels as alphabet. 
For example $P^*$ matches paths with zero or more edges labelled $P$, $P?$ matches paths with zero or one $P$ label, and $\epsilon$ the empty path.
Edges $s_1 \xrightarrow{P} s_2$ are used to denote that $s_2$ is the parent scope of $s_1$.
Due to the structure of XMI, $\mathcal{G}$ is structured as a tree with respect to label $P$ such that from every scope $s$, root node $s_r$ is reachable using path $P^*$.
Notation $\nabla s$ is used to indicate a fresh scope not part of $\mathcal{G}$.

\subsection{Abstract notation for XMI}
The SysML specification can be exported to XML Metadata Interchange (XMI) version 2.1 \cite{omg2005xml}.
This textual and computer-interpretable format is used as input to our transformation framework and parsed by Spoofax to an AST.

\begin{figure}[!ht]
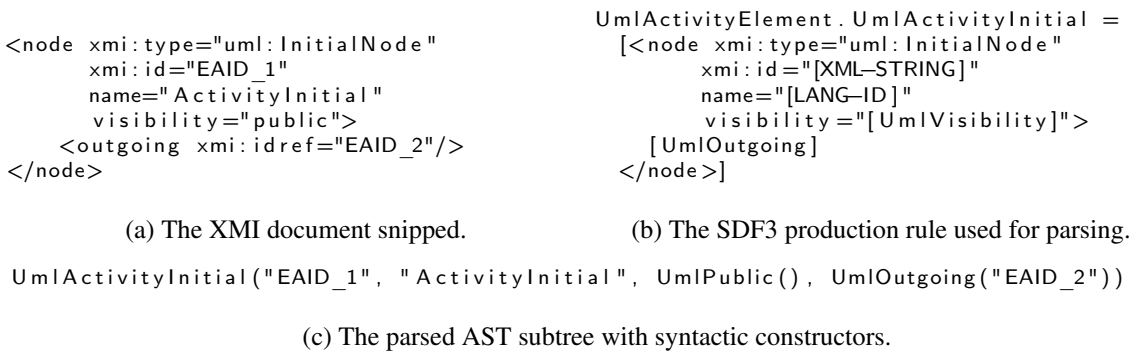

  \begin{subfigure}{0.48\textwidth}
    \centering
    \vspace{2ex}
    \begin{lstlisting}
<node xmi:type="uml:InitialNode"
      xmi:id="EAID_1"
      name="ActivityInitial"
      visibility="public">
    <outgoing xmi:idref="EAID_2"/>
</node>
    \end{lstlisting}
    \caption{The XMI document snipped.}
  \end{subfigure}
  \begin{subfigure}{0.48\textwidth}
    \centering
    \begin{lstlisting}
UmlActivityElement.UmlActivityInitial =
  [<node xmi:type="uml:InitialNode"
        xmi:id="[XML-STRING]"
        name="[LANG-ID]"
        visibility="[UmlVisibility]">
    [UmlOutgoing]
  </node>]
    \end{lstlisting}
    \caption{The SDF3 production rule used for parsing.}
  \end{subfigure}
  \\
  \begin{subfigure}{0.98\textwidth}
    \centering
    \begin{lstlisting}
UmlActivityInitial("EAID_1", "ActivityInitial", UmlPublic(), UmlOutgoing("EAID_2"))
    \end{lstlisting}
    \caption{The parsed AST subtree with syntactic constructors.}
  \end{subfigure}
  \caption{AST construction from XMI document in Spoofax.}
  \label{fig:spx-production}
\end{figure}

To represent the parsed document in the AST, Spoofax generates syntactical constructor type definitions based on the SDF3 syntax specification.
These syntax specifications can be specified using string templates and non-terminal symbols.
Consider the representation of an Activity Initial, a start point of the control flow of an AD, in an XMI document.
Assuming that the referenced symbols between [ and ] brackets are defined, a SDF3 production rule for symbol UmlActivityElement and constructor name UmlActivityInitial is given in Figure \ref{fig:spx-production}.

In the remainder of this paper, abstract versions of these syntactic constructors of 
the parsed XMI format are used to describe the transformation and semantical elements textually.
Some SysML constructs are mapped to the same XMI elements. 
All relevant SysML elements with their syntactic constructor notation are listed in 
Table \ref{table:xmi} where each element gets a unique
ID $i$.
In flows the directed edge relations in the ADs are drawn dashed for the control 
flows and solid for object flows. These are used to determine the flow type. 

\begin{table}[ht!]
  \centering
  \SetTblrInner{rowsep=3pt,colsep=3pt}
  \begin{tblr}{colspec={m{0.240\linewidth} | m{0.5\linewidth} | m{0.195\linewidth}}}
    \textbf{Abstract constructor}        & \textbf{Constructor parameter description} & \textbf{Visualisation} \\
    \hline
    $\text{EnumerationLiteral}(i,n)$     & ID $i$,
    name $n$                             & \SetCell[r=2]{c}{
      \includegraphics[height=3em]{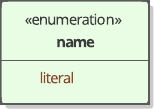}
    }
    \\
    \cline{1-2}
    $\text{Enumeration}(i,n,L)$          & ID $i$,
    name $n$,
    set of enumeration literals $L$      &
    \\
    \hline
    $\text{Property}(i,n,t)$             & ID $i$,
    name $n$,
    ID reference of property type $t$    & \SetCell[r=2]{c}{
      \includegraphics[height=3em]{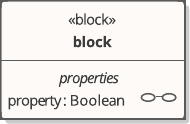}
    }
    \\
    \cline{1-2}
    $\text{Block}(i,n,P)$                & ID $i$,
    name $n$,
    set of block properties $P$          &
    \\
    \hline
    $\text{Attribute}(i,n,t,d)$          & ID $i$,
    name $n$,
    ID reference of attribute type $t$,
    default value $d$ of referenced type & \SetCell[r=1]{c}{
      \includegraphics[height=2em]{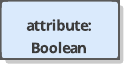}
    }
    \\
    \hline
    $\text{ActivityParameter}(i,n,t)$    & ID $i$,
    name $n$,
    ID reference of parameter type $t$   & \SetCell[r=1]{c}{
      \includegraphics[height=2em]{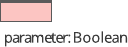}
    }
    \\
    \hline
    $\text{ActivityInitial}(i)$          & ID $i$
                                         & \SetCell[r=3]{c}{
      \includegraphics[height=2em]{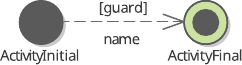}
    }
    \\
    \cline{1-2}
    $\text{ActivityFinal}(i)$            & ID $i$
    \\
    \cline{1-2}
    $\text{Flow}(i,f,n,g,s,t)$           & ID $i$,
    flow type $f \in \lbrace \text{control}, \text{object}\rbrace$,
    name $n$ or empty name $\epsilon$,
    guard $g$ or empty guard $\delta$,
    source ID reference $s$,
    target ID reference $t$
    \\
    \hline
    $\text{Pin}(i,n,t)$                  & ID $i$,
    name $n$,
    ID reference of parameter type $t$
                                         & \SetCell[r=2]{c}{
      \includegraphics[height=2em]{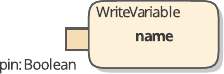}
    }
    \\
    \cline{1-2}
    $\text{WriteVariable}(i,n,p)$        & ID $i$,
    variable name $n$,
    input pin $p$
    \\
    \hline
    $\text{CallBehaviour}(i,n,b,P)$      & ID $i$,
    name $n$,
    ID reference of behaviour $b$,
    set of parameters $P$
                                         & \SetCell[r=1]{c}{
      \includegraphics[height=2em]{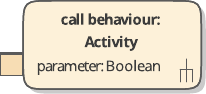}
    }
    \\
    \hline
    $\text{DecisionNode}(i,n)$           & ID $i$,
    name $n$
                                         & \SetCell[r=1]{c}{
      \includegraphics[height=2em]{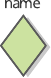}
    }
    \\
  \end{tblr}
  \caption{Table with abstract syntactic constructors for SysML XMI elements.}
  \label{table:xmi}
\end{table}

\subsection{Well-typedness using constraint solving in Statix}
The SysML models are structured in a tree of packages.
The root of the XMI document contains the root package.
The packages help to guide humans in navigating the model.
The document is well-typed if and only if the root is well-typed,
and the root, or for that matter any node, is well-typed iff all its children are well-typed,
although for each node extra typing constraints may be required.

\begin{figure}[!ht]
  \begin{subfigure}{.49\textwidth}
    \centering
    \includegraphics[width=.98\linewidth]{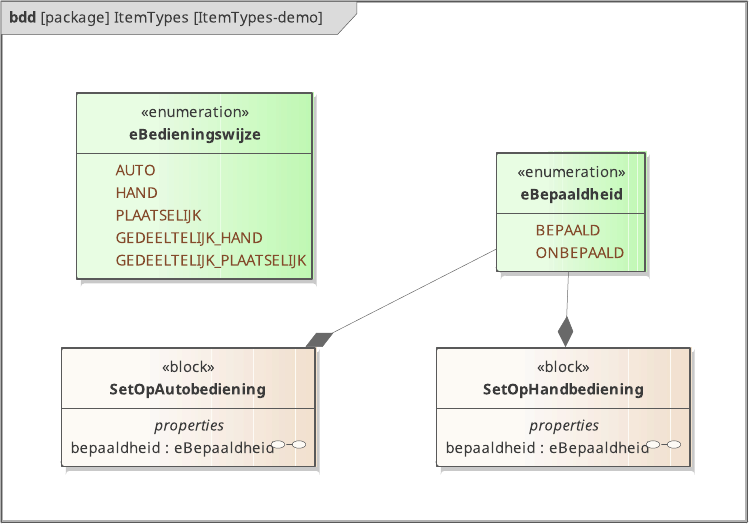}
    \caption{The SysML BDD snippet.}
    \label{fig:itemtypes-bdd}
  \end{subfigure}
  \begin{subfigure}{.49\textwidth}
    \centering
    \includegraphics[width=.98\linewidth]{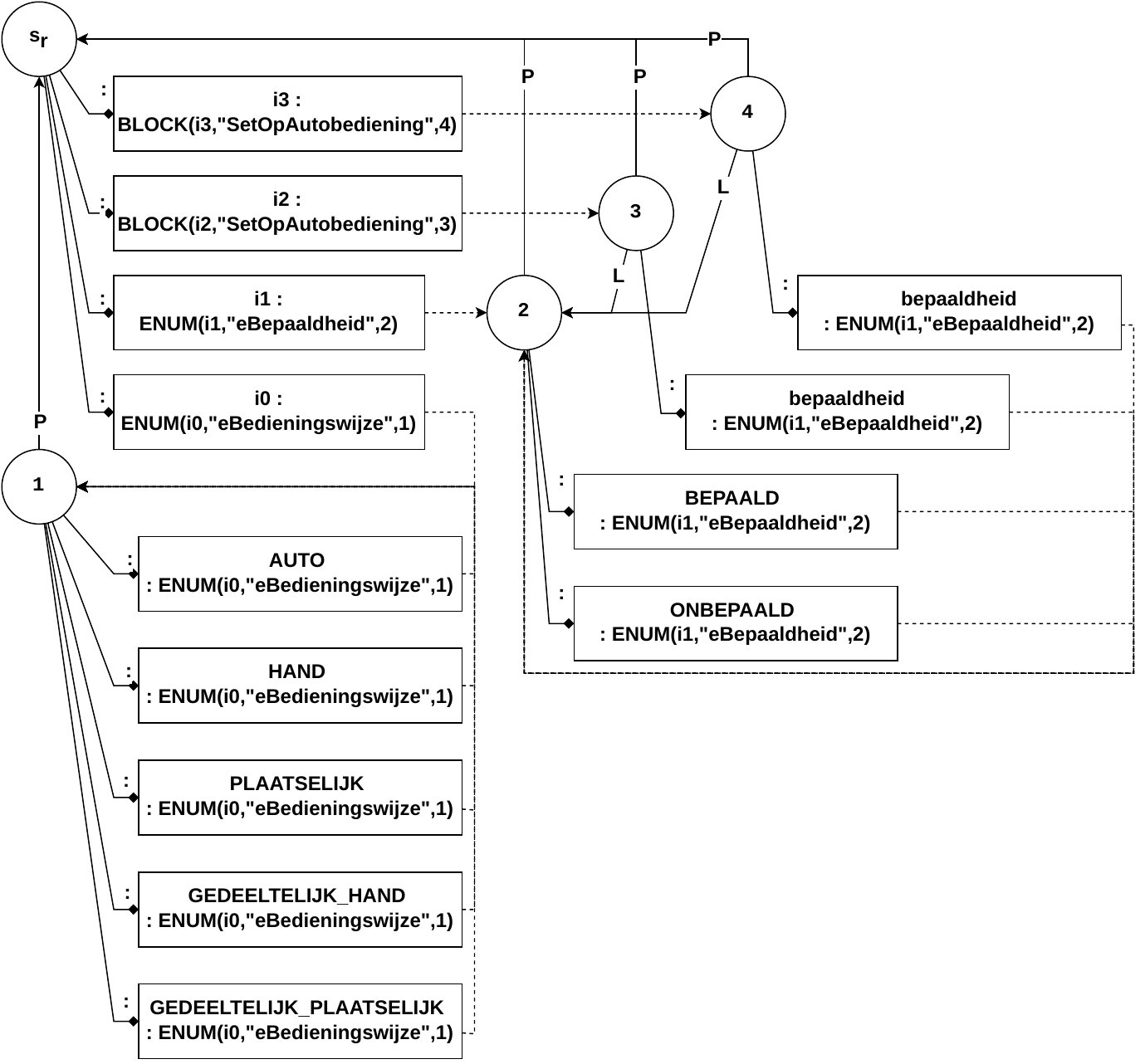}
    \caption{The corresponding scope graph of Figure \ref{fig:itemtypes-bdd}.}
    \label{fig:itemtypes-sg}
  \end{subfigure}
  \caption{A snippet of enum and block definitions in a Block Definition Diagram (BDD) of the model.}
  \label{fig:itemtypes}
\end{figure}

These extra typing constraints are formulated in Statix using inference rules. As an example we give the rule for
an enumeration that typically belongs to an enumeration declaration as in Figure \ref{fig:itemtypes-bdd}.

\begin{align*}
  \frac
  {\nabla s_e \quad s_e \xrightarrow{P} s_r \quad T \equiv \text{ENUM}(i,n,s_e) \quad s_r \dat{\colon} (i,T) \quad \forall_{\text{EnumerationLiteral}(i_l,n_l) \in L}\ {s_e \dat{\colon} (n_l, T)}}
  {\text{Enumeration}(i,n,L)_\ok}
\end{align*}

In order to understand this rule it is important to know that these rules use scope graphs,
see for instance Figure \ref{fig:itemtypes-sg}. A scope graph is a directed graph of scopes with labelled links
in which objects
are declared. By searching the scope graph the nearest scope can be found in which an object is declared to
determine its type and other properties. 
In the rule above $\nabla s_e$ says that a new scope $s_e$ is added to the scope graph,
and $s_e \xrightarrow{P} s_r$ says that this new scope is linked the existing scope $s_r$. 

The rule above now expresses that an enumeration is well-typed, $\text{Enumeration}(i,n,L)_\ok$,
if the enumeration with unique identifier $i$ is added to scope $s_r$ with type $T=\text{ENUM}(i,n,s_e)$.
Furthermore, all elements $n_l$ of this enumeration, which occur in list $L$, are added with the same type
$T$ to the new scope $s_e$. 

More examples of well-typedness rules are given in  Appendix \ref{app:spoofax}.
Besides the links labelled with $P$ for name resolution, additional edges in the scope graph occur
labelled with $T$ and $L$ to connect the scope of the block to the scopes of the 
referenced semantic type BLOCK and ENUM respectively.
Figure \ref{fig:itemtypes} shows an example snippet of enum and block definitions, together with the constructed 
scope graph.
Visually, referenced scopes in semantic type constructors are connected with a dashed line.
For the translation to mCRL2, every declaration of semantical type ENUM or BLOCK in $s_r$ is directly translated 
to a sort specification.

In the next sections, AD elements that represent typed values are declared as variables of the referenced type.
As semantic types are declared using their unique XMI ID in the root scope, they can always be resolved by following a $P^{*}$ path from the scope to $s_r$.
Flows that connect elements in the diagram also connect the nodes in the scope graph with labelled edges.
References to block properties or enumeration literals can be accessed using a dot notation.
Judgement $s \vdash e \colon T$ denotes that in the context of scope $s$, expression $e$ has type $T$. 
Judgement $s \vdash p \xrightarrowdash{r} d$ states that data term $d$ is visible through path query $p$ from 
scope $s$ under relation $r$.
Trivially, $s \vdash n \colon T$ if $s \vdash \epsilon \xrightarrowdash{\colon} n \colon T$.

We show the typing rules to refer to a property of a block $n_1.n_2$, 
to equality $\square \in \lbrace \equiv, \not\equiv \rbrace$,
to negation $\neg$, and to binary boolean operators
$\Delta \in \lbrace \land, \lor \rbrace$. 
\begin{align*}
  \frac
  {s \vdash n_1 \colon T_1 \quad \text{scope}(T_1) \vdash \epsilon \xrightarrowdash{\colon} n_2 \colon T_2}
  {s \vdash n_1.n_2 \colon T_2}
  &  &
  \frac
  {s \vdash n_1 \colon T \quad s \vdash n_2 \colon T}
  {s \vdash n_1\ \square\ n_2 \colon \mathbb{B}}
  &  &
  \frac
  {s \vdash n \colon \mathbb{B}}
  {s \vdash \neg n \colon \mathbb{B}}
  &  &
  \frac
  {s \vdash n_1 \colon \mathbb{B} \quad s \vdash n_2 \colon \mathbb{B}}
  {s \vdash n_1\ \Delta\ n_2 \colon \mathbb{B}}
\end{align*}

\subsection{Leaf Decomposition Layer}
The elements in Activity Diagrams representing typed values are declared in the scope graph while establishing the well-typedness of the AST elements.
To demonstrate how variables are declared and their linked types resolved, consider the activity parameters of the diagram in Figure \ref{fig:leaf-ad}.
Let $\nabla s_a$ with $s_a \xrightarrow{P} s_r$ and $s_r \dat{\colon} i \colon \text{ACT}(i,s_a)$ be the scope and semantic type declaration of the AD with id $i$.
For every such parameter $\text{ActivityParameter}(i,n,t)$, the well-typedness conclusion is established, making sure the referenced type exists, and a declaration of the semantic type PARAM in scope $s_a$ is created.

\begin{align*}
  \frac
  {\nabla s_p \quad s_p \xrightarrow{P} s_a \quad s_p \xrightarrow{T} \mathit{scope}(T) \quad s_p \vdash P^* \xrightarrowdash{\colon} t \colon T \quad s_p \xrightarrow{\colon} n \colon T \quad s_a \dat{\colon} (i,\text{PARAM}(i,n,T,s_p))}
  {\text{ActivityParameter}(i,n,t)_\ok}
\end{align*}

The well-typedness establishment of an object flow from $s$ to $t$ connects the flow scope $\nabla s_f$ with $s_f \xrightarrow{E} \mathit{scope}(s)$.
Other derivations make sure to finish the path from the target to scope $s_f$ by inserting $\mathit{scope}(t) \xrightarrow{E} s_f$.
It is sufficient to only allow if-then-else decision nodes, as this trivial split in decisions makes the diagrams surprisingly easy to read for modellers whilst being sufficiently expressive.
Every decision node has either one outgoing object flow with empty guard $\delta$, or one flow with as guard special value $\texttt{else}$ flowing to the next decision node and one object flow with a guarded value assignment $e_1 := e_2$ as name $n$ to an output activity parameter.
Value $e_2$ must be visible by a path following the object flows in reverse direction with $E^*$.
To resolve enumeration literals as constants for enumerations in scope, $T^+ L$ is added to the allowed paths.

\begin{align*}
  \frac
  {\mathit{scope}(t) \vdash \epsilon \xrightarrowdash{\colon} e_1 \colon T \quad \mathit{scope}(s) \vdash E^*(T^+ L)? \xrightarrowdash{\colon} e_2 \colon T}
  {(e_1 := e_2)_\ok}
\end{align*}

\begin{figure}[!ht]
  \begin{subfigure}{.49\textwidth}
    \centering
        \vspace{4ex}
    \includegraphics[width=.98\linewidth]{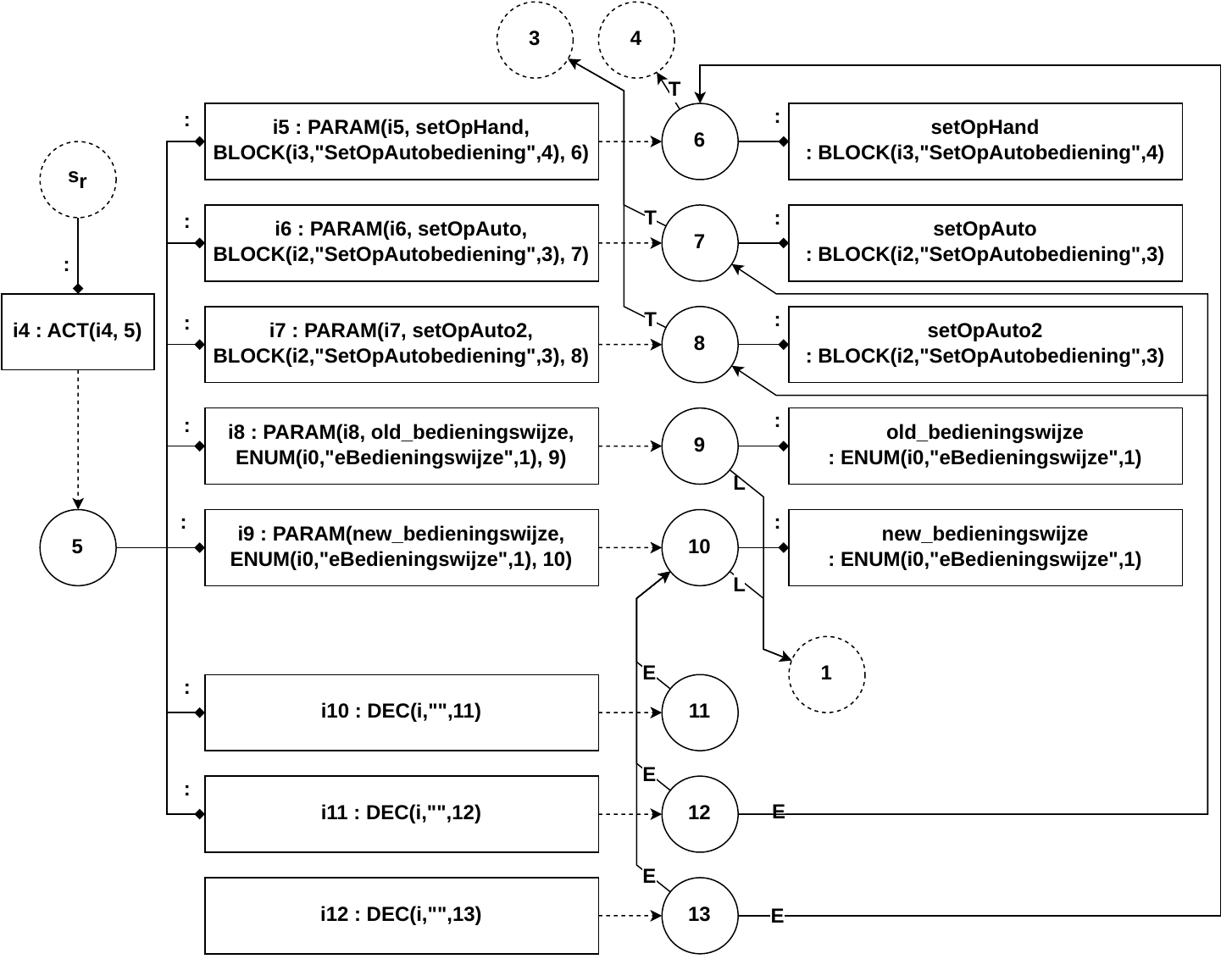}
    \vspace{2ex}
    \caption{The extension of Figure \ref{fig:itemtypes-sg} excluding $\xrightarrow{P}$ edges.}
    \label{fig:leaf-sg}
  \end{subfigure}
  \begin{subfigure}{.49\textwidth}
    \centering
    \begin{lstlisting}
map
  compute_bedieningswijze2 : eBedieningswijze # SetOpAutobediening # SetOpAutobediening # SetOpHandbediening -> eBedieningswijze;
var
  old_bedieningswijze2 : eBedieningswijze, setOpAuto : SetOpAutobediening, setOpAuto2 : SetOpAutobediening, setOpHand : SetOpHandbediening;
eqn
  compute_bedieningswijze2(old_bedieningswijze2, setOpAuto, setOpAuto2, setOpHand) = if ((bepaaldheid5(setOpHand) == BEPAALD), HAND, if (((bepaaldheid4(setOpAuto) == BEPAALD) || (bepaaldheid4(setOpAuto2) == BEPAALD)), AUTO, old_bedieningswijze2));
    \end{lstlisting}
    \caption{The generated mCRL2 specification.}
    \label{fig:leaf-mcrl2}
  \end{subfigure}
  \caption{Generated scope graph and mCRL2 specification of Figure \ref{fig:leaf-ad}.}
  \label{fig:leaf}
\end{figure}

An important benefit of these very restrictive semantics is that the diagrams describe a completely deterministic computation procedure.
As a result, the diagrams can be completely expressed using data expressions in mCRL2 without the need to encode the semantical rules of the diagram in the resulting process specification.
To ease the binding of input and output variables in mCRL2, a map declaration is defined for every diagram as demonstrated in Figure \ref{fig:leaf-mcrl2}.

\subsection{Glue Decomposition Layer}
The ADs in the glue layer use the same definitions for the flows in the diagrams.
Decision nodes are not supported in this layer.
The behaviour calls are calling the ADs defined in the previous section.
In the translation of the AD to an mCRL2 process specification, all behaviour calls are translated as $\sum_{o \colon T_o} o = \text{ActivityMapping}(in_1,\ldots, in_n)$ where $o$ is a fresh variable name of output type $T_o$, which is set equal to the result of calling the mapping with the bound input parameters $in_1$,\ldots, $in_n$.
This process specification follows the pattern that can be optimised by sum elimination in the mCRL2 toolset.
The attributes are translated to process parameters in the process equation representing the AD.
All WriteVariable elements are translated to process variable updates in the recursive process specification.
Again, the very restrictive subset of diagram elements permitted in this layer makes it trivial to express the semantics of every diagram by a process specification in mCRL2 by defining a specification with state variables for every attribute, 
using the previously defined maps and recursing in itself with updated state variables using the WriteVariable calls.
Figure \ref{fig:glue-ad} depicts such a glue AD and the generated mCRL2 is moved to Appendix \ref{app:glue-mcrl2} due to its size.

\subsection{Strengths and weaknesses}
The translation sketched above has the following strengths.
It is compositional and can be applied to translate the complete SysML 
tunnel specification by putting all components in parallel and combining the
synchronous input and output into multi-actions in mCRL2 as explained in
\cite{jilissen2023formal}.

All model elements added to the approach are understandable with the SysML knowledge that was already required to 
understand the previous model.
In this way, all engineers at RWS and its contractors are able to read and specify these kinds of diagrams.
Another strong reason to choose this approach is that with our extension
to replace `structured natural language' the complete tunnel
specification can be formulated as a large coherent and reasonably precise 
SysML specification. 

Unfortunately, the SysML tooling does currently not enforce that the SysML
specification is well-typed and internally consistent.
But this is remedied as the static analysis using Statix of the exported XMI 
type checks the model and gives feedback on mistakes.
The value of the defined semantics for the AD layers is shown in an analysis of 
the generated mCRL2 specification.
Even though the presented work mainly focusses on the formal modelling and specification, all verification results of the previously analysed properties of the manual translation \cite{jilissen2022formal} could be replicated on the generated specification, as shown in Table \ref{table:verification}.
This is not an exhaustive list of all properties that must be verified on the system, but rather a selective subset which demonstrates that several types of properties are feasible to verify.
The properties are checked for the complete sub-system instantiated with one instance of the BF, SF, and SP.
The replication also includes verifying the modification of the model introduced in \cite{jilissen2022formal} to fix the model such that a bumpless transfer is guaranteed.
The $\mu$-calculus formulae with adapted naming to fit the generated model are included in the model repository of this paper.

\begin{table}[!ht]
  \centering
  \begin{tabular}{c|m{0.61\linewidth}|c|c}
    \textbf{Requirement} & \textbf{Description} & \textbf{H} & \textbf{S} \\
    \hline
    Deadlock freeness & The control system can never reach a state in which no progress is possible. & \checkmark & \checkmark \\
    \hline
    Configured manual stand & The control system only sends the last configured manual-control values to the controlled installation when the control mechanism is manual-control. & \checkmark & \checkmark \\
    \hline
    No spontaneous change & The control system can never change the value sent to the controlled installation without explicitly receiving commands to do so. & \checkmark & \checkmark \\
    \hline
    Bumpless transfer & The control mechanism must be altered `bumpless': there should not be an immediate change in behaviour of the control system without explicit requests to do so. & $\times$ & $\times$ \\
  \end{tabular}
  \caption{Verification results of the hand-written (H) and SysML (S) mCRL2 translation.}
  \label{table:verification}
\end{table}

The main weakness of this SysML based style is that every detail has to be specified graphically.
To create the diagrams for this model, a lot of manual labour is needed to draw all the relations with components and to specify all flow names using the assignment language, as can already been seen by the glue layer of the BF in Figure \ref{fig:glue-ad}.
Assigning and aligning all flows and flow name labels is sheer drudgery, and even with automated routing of flows the readability suffers greatly.
With the huge number of diagrams, it is almost impossible for an engineer to keep
an overview of the whole system. 
It is therefore questionable whether graphical formalisms such as SysML are
fundamentally the most efficient way to model system with the complexity of tunnel
control systems. 

Another important weakness of this approach is the style in ADs to let 
all inputs and outputs occur simultaneously. Although, this matches neatly on multi-actions in mCRL2, it fundamentally leads to an exponential growth in the 
number of outgoing transitions in each state of the mCRL2 model. This
puts fundamental constraints on the analysability of the whole tunnel control model.

\section{Alternative models using Dezyne}
As the SysML models are not ideal, it has been investigated whether 
the Dezyne specification language \cite{beusekom2021dezyne} offers a viable alternative.
Dezyne offers a syntax that looks similar to widely used programming languages in industry and has
visualisations closely related to SysML, whilst fitting in the current modelling workflow. 
Therefore, the adoption of the language within RWS is deemed feasible.

In Dezyne, components are defined which communicate over ports of formally specified interfaces with observable behaviour.
Components can be composed together in systems by connecting ports with ports of other components or 
letting them communicate with ports of the environment.
Dezyne requires that a number of properties are verified before generating code,
such as freedom of deadlock, absence of illegal behaviour and interface compliancy meaning that a component exactly provides 
the interface that had to be separately specified.
Under the hood, Dezyne uses mCRL2 as its verification engine and can export specifications and transition systems in formats usable in the mCRL2 tooling \cite{beusekom2017formalising}.
%In Dezyne, an execution model is used that enforces single-threaded run-to-completion semantics in the system \cite{beusekom2017formalising}.
%Even though this language has a completely different runtime semantics than the encoding created for SysML in mCRL2, it can give insights in the specified behaviour.

As a proof-of-concept, the SysML specification of the overpressure controller is manually translated to Dezyne.
In the SysML model, all inputs arrive together in one big event. 
In the translation, such big events are
split-up by letting each part of the input employ a single command or status event as Dezyne is not designed nor intended to handle
such massive input events. 

In the SysML model the massive input contains fields to indicate whether a particular input is to be considered to be 
present or absent in the input. 
In Dezyne this information is not necessary as input events with data that is absent simply do not need to take place. 
Omitting the indicator that data is present or absent has a hugely reducing effect on the complexity of the state space.

As illustration, consider the interface 
combining the two commands blocks in Figure \ref{fig:itemtypes-bdd}. This interface is translated to Dezyne as an 
interface containing two in-events SetOpAutobediening and SetOpHandbediening without parameters.

An issue with Dezyne version 2.18, is that values in the input cannot be used within the model and can only be passed on.
This means that values in the input that have an influence on the behaviour are encoded in the event name. 
Unfortunately, this applies to all enumeration values in the investigated models. The reason for this design choice in Dezyne
is that data that influences the behaviour substantially increases the size of state spaces, and hence hampers the possibility
to verify properties.

Using these interface definitions, every AD of the original SysML specification is modelled as a component 
in Dezyne.  We used two different styles of modelling.
In the first style, components query the information they need from other components on-demand to determine their response to events.
We call this the pull style. 
In the second style, components push all information that other components might need to these components whenever it becomes available.

\subsection{The pull style model}
When using pull style models, the components query their required interfaces for all information they need to handle incoming events.
The main benefit of this modelling approach is that it does not require introducing shadow variables to store the last received state information in multiple components.
This prevents a large growth in the state space \cite{groote2015specification} caused by copies of values which do not atomically 
change due to the delay in propagation throughout the system.
To achieve this, instead of having parameterised in-events to communicate the value of some property, an out-event is defined to indicate that the component interacting over the interface wants to know the value.
In response to this event, the communicating component must reply with any of the enumeration literals permitted by the interface definition.
With this approach, it is possible to generate the complete state space of 
each component and analyse those using the formal analysis in mCRL2. A typical trace is given in 
Figure \ref{fig:dzn-pull}. Under substantial abstractions, we could even show that this model was bisimilar to the SysML
model. 

The main downside of this approach is that it does not benefit of most of the verification features offered by Dezyne itself.
In the pull style, we need to model interfaces mainly in a stateless way. 
This has as a consequence that the verification methods that Dezyne offers
such as the compliance verification of the implementations in components 
largely loose their value. Verification of the system as a whole is also
troublesome as generating the overall behaviour from the components is 
currently not feasible. 

\subsection{The push style model}
To make more use of the verification capabilities and align with the reactive 
components principles of Dezyne, the system has also been modelled in a push style \cite{groote2015specification}. 
In this style all components forward information to other components as soon as it becomes available.
This approach makes it possible to define interfaces with meaningful states and rely on the verification 
of the Dezyne tooling. We modelled the overpressure system in the same style as in SysML, using `diamond 
patterns' where subcomponents are controlled by multiple super-components, which is a style that
Dezyne does not encourage. This forces us to use so called 
\textit{optional trigger events} to push information to various super-components. 
Unfortunately, this causes the interfaces to grow, which cannot be handled by the verification engine of Dezyne.
Even when the Dezyne tooling is executed on a large server with 3TB of memory, the verification of interfaces and component compliance with these interfaces takes days or does not finish within a week.

\begin{figure}[!ht]
  \begin{subfigure}{.49\textwidth}
    \centering
    ~\vspace{4ex}\\
    \includegraphics[width=\textwidth]{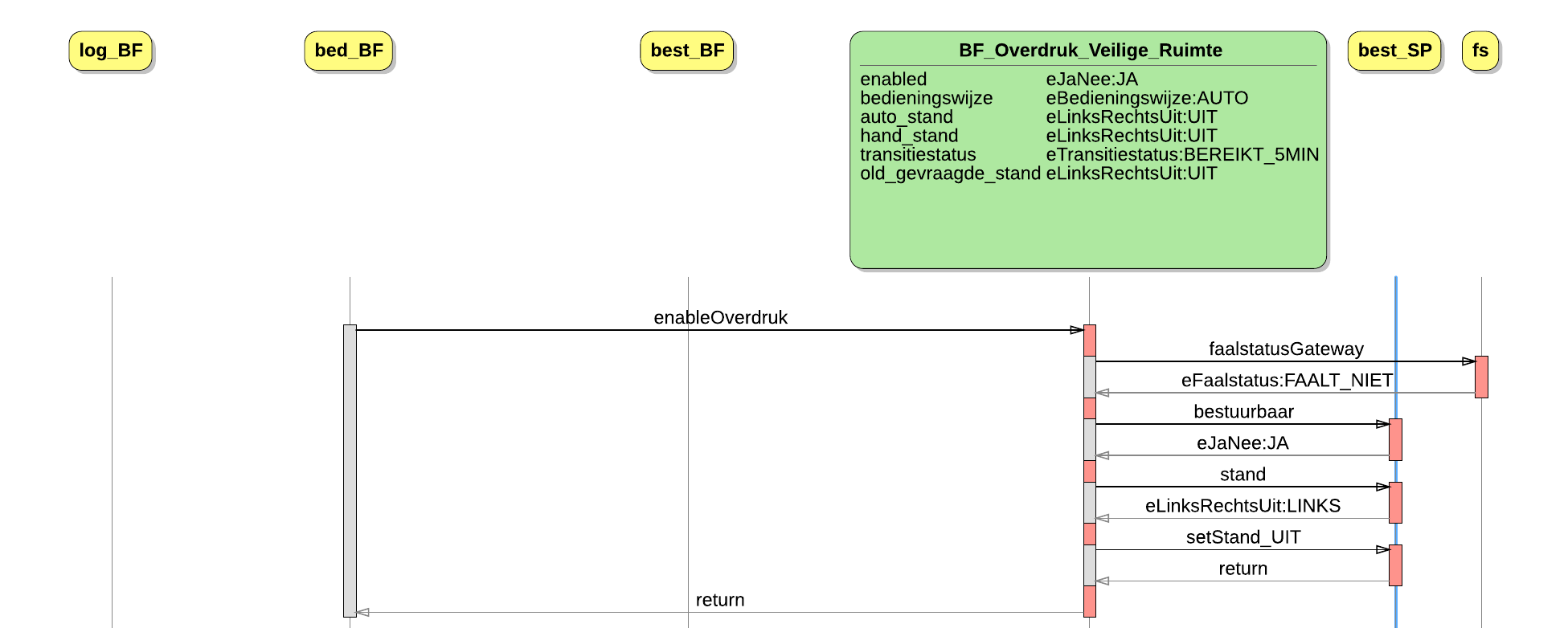}\\
    ~\\
    \caption{A trace of the overpressure system in the pull model.}
    \label{fig:dzn-pull}
  \end{subfigure}
  \begin{subfigure}{.49\textwidth}
    \centering
    \includegraphics[width=\textwidth]{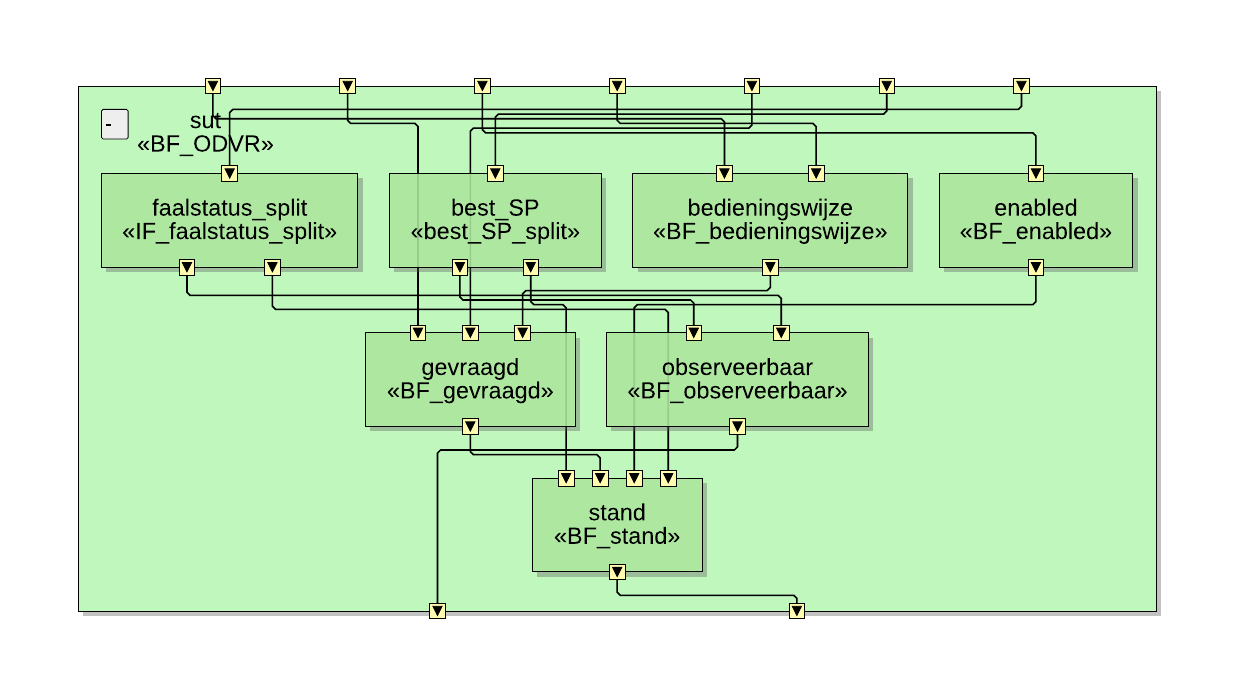}
    \caption{Decomposition into many computation components.}
    \label{fig:dzn-decomposition}
  \end{subfigure}
  \caption{Model visualisations of Dezyne.}
\end{figure}

We remedy this situation by 
decomposing the system into many more components which compute standalone values 
providing and requiring much smaller interfaces. This is illustrated in Figure \ref{fig:dzn-decomposition}.
With this decomposition and abstraction, Dezyne is able to verify its standard properties on all components and interfaces.

%A more scalable solution is a similar decomposition extension as introduced for SysML in Section \ref{sec:assignment} can be applied, as 
%Instead of modelling as Dezyne components, a system can be defined composed of the newly introduced decomposition 
%In this decomposition, additional abstractions have been applied to reduce the state space of the interfaces and components.
%For example, the computation of the observability of a component relies on the absence of several error statuses.
%These statuses can be combined on an as low as possible level to prevent having multiple boolean values being passed around.

\subsection{Verification of general properties}
Applying the built-in verification techniques to Dezyne models is known to have a very beneficial effect
on the quality of the models both in development time and number of faults \cite{DBLP:conf/icsm/GrooteOW11}.
However, we fail to verify those properties over the whole state space that were so useful in increasing the
primary quality of the tunnel model in \cite{jilissen2023formal}. 
In the pull style it was impossible to generate the overall state space of the system although the Dezyne
toolset provides some means to do so. In push styles
state spaces will become much bigger \cite{groote2015specification}. 
Although we feel that the Dezyne toolset could do a better job in state space generation, 
we do not expect this to be available for quite some years to come,
especially because the way Dezyne generates a full state space does not employ
symbolic techniques or parallelism \cite{groote2022using,DBLP:conf/tacas/LaveauxWW22}.

\section{Conclusions}
In this paper, the promising results in \cite{jilissen2023formal} are being elaborated to come
to an industrially viable environment in which existing SysML models of road tunnel control systems
can be analysed. 
In the first approach, a restrictive extension to the SysML specification is introduced which precisely 
specifies details with respect to the behaviour of the system which were either written down in natural language or were completely absent. 
While this approach allows the generation of the complete state space of sub-systems and formal verification of 
both liveness and safety properties over these systems, two major problems became obvious. The first one
is that the modelling style of RWS leads to transition systems with massive fanout hampering verification when systems
become large. More worrying is that the SysML modelling style is leading to an unwieldy cluttering of graphical
entities which is very time consuming to make, and utterly
hard for humans to keep track of. 

As an alternative the commercial modelling language Dezyne has been investigated.
This language offers benefits because it verifies a wide range of properties of the 
models that have been shown to substantially improve the quality of the specification.
But these properties are primarily verified on individual components, and as it stands
it does not appear to be easily possible to systematically verify the global properties
of tunnel control systems. Although we believe that languages such as Dezyne can and will
ultimately be developed further, they are as it stand insufficient as a work horse for tunnel
control model. 

This opens up the question on the next step within organisations such as Rijkswaterstaat as the possibility to verify 
has clearly shown itself indispensable in \cite{jilissen2023formal}. 
As it stands we believe that proceeding with commercially available tools is not the most profitable way to go. 
Academically developed 
specification languages such as LOTOS NT \cite{garavel2013cadp} and mCRL2 \cite{bunte2019mcrl2}
might be more suitable as they have flexible and very expressive formalisms to express behaviour and correctness properties,
and are actually supported by stable and capable toolsets that are available and reliable for decades. 
The interesting question is whether these academically developed languages will be picked up as primary tools
within the context of for instance Rijkswaterstaat. 

\bibliographystyle{eptcs}
\bibliography{generic}

\appendix

\newpage
\section{Artefacts}
\label{app:artefacts}
The artefacts belonging to this paper are available for download at \url{http://mars-workshop.org/repository.html}.
Further instructions are given on the repository page for this paper.

\newpage
\section{Generated mCRL2 of the Activity Diagram in Figure \ref{fig:glue-ad}}
\label{app:glue-mcrl2}
In Figure \ref{fig:glue-ad}, the Activity Diagram of the glue layer of the Base Functionality of the Overpressure Safe Space sub-system is shown.
This diagram is automatically translated to the mCRL2 process specification below.
In this specification, the mapping names and equations starting with \texttt{compute_} are generated from the ADs which are called by the Call Behaviour SysML diagram elements.
The binding of variables is defined by the assignment in the name of object flows in the glue AD.

\begin{lstlisting}
proc BF_Overdruk_Veilige_Ruimte(
  transitiestatus2 : eTransitiestatus,
  hand_stand : eLinksRechtsUit,
  enabled3 : eJaNee,
  bedieningswijze4 : eBedieningswijze,
  auto_stand : eLinksRechtsUit
) = sum
  best_BF2 : best_Coordinatie_Luchtkwaliteit_Veilige_Ruimte, 
  faalstatus2 : MonitoringIntegriteit3BBesturingssysteemVeiligeRuimte_FaalstatusGateway,
  sp_tk2 : besttk_BF_Overdruk_Veilige_Ruimte_SP_Overdruk_Veilige_Ruimte,
  bed_BF : bed_BF_Overdruk_Veilige_Ruimte,
  new_stand3 : eLinksRechtsUit,
  new_stand2 : eLinksRechtsUit, 
  setStand2 : SetStand,
  new_gevraagde_stand : eLinksRechtsUit,
  new_bedieningswijze2 : eBedieningswijze,
  observeerbaar4 : eJaNee,
  beschikbaarheid3 : eBeschikbaarheid,
  disabled5 : eJaNee,
  new_enabled : eJaNee,
  new_transitiestatus : eTransitiestatus,
  old_gevraagde_stand : eLinksRechtsUit,
  statusMtkLuiken2 : eOpenDicht
. (
  statusMtkLuiken2 == compute_statusmtkluiken(
    observeerbaar4,
    luikenGesloten2(sp_tk2))
  && old_gevraagde_stand == compute_old_gevraagde_stand(
    auto_stand,
    bedieningswijze4,
    hand_stand)
  && new_transitiestatus == compute_transitiestatus(
    new_gevraagde_stand,
    old_gevraagde_stand,
    transitiestatus2,
    stand5(sp_tk2))
  && new_enabled == compute_enabled2(
    disableOverdruk(bed_BF),
    enableOverdruk(bed_BF),
    enabled3)
  && disabled5 == compute_disabled3(new_enabled)
  && beschikbaarheid3 == compute_beschikbaarheid2(
    bestuurbaar3(sp_tk2),
    faalstatusGateway(faalstatus2),
    storingOverdrukregeling2(sp_tk2),
    storingOverdrukventilatoren2(sp_tk2),
    new_transitiestatus)
  && observeerbaar4 == compute_observeerbaar2(
    bestuurbaar3(sp_tk2),
    faalstatusGateway(faalstatus2),
    redenNietBestuurbaarOpstart(sp_tk2),
    redenNietBestuurbaarStoring3(sp_tk2))
  && new_bedieningswijze2 == compute_bedieningswijze2(
    bedieningswijze4,
    setOpAutobediening2(bed_BF),
    setOpAutobediening(best_BF2),
    setOpHandbediening(bed_BF))
  && new_gevraagde_stand == compute_gevraagde_stand(
    new_stand2,
    new_bedieningswijze2,
    new_stand3)
  && setStand2 == compute_setstand(
    bestuurbaar3(sp_tk2),
    new_enabled,
    faalstatusGateway(faalstatus2),
    new_gevraagde_stand,
    stand5(sp_tk2))
  && new_stand2 == compute_auto_stand(
    auto_stand,
    setAutobedieningsStand(best_BF2))
  && new_stand3 == compute_hand_stand(
    new_bedieningswijze2,
    hand_stand,
    setHandbedieningsStand(bed_BF))
  && true
) -> 
  bedtk_BF(bedtk_BF_Overdruk_Veilige_Ruimte(
    new_stand2,
    new_bedieningswijze2,
    beschikbaarheid3,
    bestuurbaar3(sp_tk2),
    disabled5,
    new_enabled,
    new_gevraagde_stand,
    new_stand2,
    luikenGesloten2(sp_tk2),
    observeerbaar4,
    redenNietBestuurbaarPlaatselijkeBediening3(sp_tk2),
    redenNietBestuurbaarStoring3(sp_tk2),
    stand5(sp_tk2),
    statusMtkLuiken2,
    storingAlgemeen3(sp_tk2),
    storingCommunicatieUitgevallen3(sp_tk2),
    GEEN_STORING,
    storingOverdrukregeling2(sp_tk2),
    storingOverdrukventilatoren2(sp_tk2),
    new_transitiestatus
  ))
  | best_sp(best_BF_Overdruk_Veilige_Ruimte_SP_Overdruk_Veilige_Ruimte(setStand2))
  | bed_BF(bed_BF)
  | sp_tk2(sp_tk2)
  | faalstatus2(faalstatus2)
  | best_BF2(best_BF2)
. BF_Overdruk_Veilige_Ruimte(
  enabled3 = new_enabled,
  bedieningswijze4 = new_bedieningswijze2,
  transitiestatus2 = new_transitiestatus,
  hand_stand = new_stand3,
  auto_stand = new_stand2
);
\end{lstlisting}

\newpage
\section{Implementations in Spoofax}
\label{app:spoofax}

As introduced in the paper, the implementation in Spoofax consists of syntax definitions, static analysis rules, and transformation rules.
This appendix elaborates on noteworthy details of the implementation and further elaborates on some concepts introduced in the paper.
The complete source code is available in Appendix \ref{app:artefacts}.

In the implementation, there are three (sub-)syntax definitions in Spoofax.
Firstly, there is the definition of the XMI language which is parsed as an Abstract Syntax Tree (AST) from the Enterprise Architect (EA) XMI 2.1 export.
Secondly, a sub-language is defined for the syntax of the Assignment-language, which is used in the syntax and semantics definition of some XMI properties.
Lastly, there is the definition of the mCRL2 language, to which the input AST is transformed, as defined by van Antwerpen et al.
\footnote{Available at \url{https://github.com/MetaBorgCube/metaborg-mcrl2/}}

\subsection{Syntax in SDF3}
For the syntax definitions, Spoofax allows defining symbols using lexical syntax definitions that match text with a regular expression.
The symbols are summarised in Table \ref{table:lexical}.

\begin{table}[h!]
  \centering
  \begin{tabular}{m{0.190\linewidth} | m{0.755\linewidth}}
    \textbf{Lexical Symbol}       & \textbf{Usage} \\
    \hline
    XML-TAG & Opening/closing tags in XML \\
    XML-PROPERTY  & Property name on XML tags \\
    XML-STRING & General-purpose property value matching some string \\
    XML-NUM & General-purpose property value matching numbers only \\
  \end{tabular}
  \caption{Table with lexical symbols for SysML XMI elements.}
  \label{table:lexical}
\end{table}

The SDF3 \cite{souza2020multi} syntax allows the definition of context-free production rules.
These production rules are string templates enclosed by [ and ] in which symbol names are enclosed in [ and ] brackets which must be further expanded.
With these definitions, parsing an XML structure could be as trivial as defining a context-free symbols \texttt{XmlProperty} and \texttt{XmlElement} with the production rules in Figure \ref{fig:spf-production}.

\begin{figure}[!ht]
  \centering
  \begin{lstlisting}
XmlProperty.XmlProperty = [[XML-PROPERTY]="[XML-STRING]"]
XmlElement.XmlElement = [<[XML-TAG][{XmlProperty " "}*]/>]
XmlElement.XmlTree = [
  <[XML-TAG][{XmlProperty ""}*]>
      [{XmlElement ""}*]
  </[XML-TAG]>
]
    \end{lstlisting}
  \caption{Production rules to parse some XML tree.}
  \label{fig:spf-production}
\end{figure}

While these rules would parse the complete XMI specification successfully, a lot of additional Statix rules would be needed to determine what the generic XML element and its properties represent and what elements are permitted in the subtree rooted at said element.
Therefore, the Spoofax language definition contains context-free symbols and production rule definitions for every item encountered in the exported XMI.
In these rules, the ordering of XML properties is fixed based on the order they appear in the XMI export for simplicity.
When not fixing this ordering, the gained flexibility would cost quite a number of Statix rules for every XMI element to verify that the required properties are given, and that exactly one of them is given.
As the exporter in EA has a stable output format, the implementation does not need this flexibility and resulting complexity.

Table \ref{table:xmi} in the paper already introduced an abstraction over the actual constructors defined in the XMI language definition.
The implementation deviates from this, for the simple reason that the XMI representation of some of the visual elements in SysML is spread out over multiple nodes in the XML structure.
Another reason for this deviation, is the ability to re-use common subtree elements in multiple locations.
As matching all these elements in a syntax definition is not hard but requires a lot of definitions, interested readers are referred to the artefacts accompanying this paper in Appendix \ref{app:artefacts}.

The sub-language for the assignments and expressions is defined.
The naming of variables in this context is defined more restrictive, to simplify the translation to mCRL2.
We opted for a lexical symbol LANG-ID which matches $[a-zA-Z\_][a-zA-Z\_0-9']^*$.
Using this symbol, references to variables are defined as:
\begin{lstlisting}
  LangVariableRef.LangVarRoot = [[LANG-ID]]
  LangVariableRef.LangVarProp = [[LangVariableRef].[LANG-ID]]
\end{lstlisting}
In this definition, we define that we can refer to variables with the LangVariableRef symbol by either directly specifying an LANG-ID or reference some property using named with a LANG-ID of some other LangVariableRef using the dot notation.

Assignments are syntactically defined as follows:
\begin{lstlisting}
  LangValue.LangVarRef = [[LangVariableRef]]
  LangAssignment.LangAss = [[LangVariableRef] := [LangValue]]
\end{lstlisting}
The LangValue symbol represents a value in the language.
As of now, only references to variables defined in the model are permitted as there are no built-in types such as booleans with pre-defined constant values.
The assignment itself can then be defined as the string template assigning a value to a variable reference.
Similar rules exist to match conditions for the guards.

\subsection{Static Semantics in Statix}
In Statix, the semantical correctness of the provided document is established by formulating a constraint problem.
As discussed in the paper, the packages serve no purpose for the semantics of the SysML diagrams.
Therefore, in the static analysis we recursively the tree of packages in the document and conclude that a package is correct if and only if we can conclude that all of its children are correct.

Consider the syntactical constructor $\text{UmlPackage}(i, n, v, C)$ as defined in Statix for packages where $i$ is the id, $n$ the name, $v$ the visibility (ignored), and $C$ the list of children.
The notation for the constraint to denote the correctness of such package becomes:

\begin{align*}
  \frac
  {\forall_{c \in \text{C}}\ c_\ok}
  {\text{UmlPackage}(i, n, v, C)_\ok}
\end{align*}

Different well-typedness constraints must be satisfied for the package elements based on the actual syntactical constructor type of the element in the AST.
In the paper, an example for the enumeration constructor was introduced.
A similar inference rule is defined below to establish the well-typedness of Blocks.
Additional edges are added to the scope graph which allow the resolution of enumeration literals in assignments and comparisons.
Two types of labels are used to facilitate this.
An edge $s_b \xrightarrow{L} s_e$ is added when declaring a property of semantical type ENUM.
For properties of semantical type BLOCK, an edge $s_b \xrightarrow{T} s_{b'}$ is added.
With the two additional inference rules below, the literals of referenced types using a dot notation can be resolved using path query $T^* L$.

\begin{align*}
  \frac
  {s_b \vdash P^* \xrightarrowdash{\colon} t_p \colon T_p \quad T_p \equiv ENUM(i_e,n_e,s_e) \quad s_b \xrightarrow{L} s_e \quad s_b \dat{\colon}(n_p, T_p)}
  {Property(i_p, n_p, t_p)_\ok}
\end{align*}
\begin{align*}
  \frac
  {s_b \vdash P^* \xrightarrowdash{\colon} t_p \colon T_p \quad T_p \equiv BLOCK(i_{b'},n_{b'},s_{b'}) \quad s_b \xrightarrow{T} s_{b'} \quad s_b \dat{\colon}(n_p, T_p)}
  {Property(i_p, n_p, t_p)_\ok}
\end{align*}

There are two inference rules to establish the well-typedness of a property.
In the first rule, the premise is a declaration $t_p \colon T_p$ where $T_p$ is of semantic type ENUM.
In the second rule, the premise is a declaration $t_p \colon T_p$ where $T_p$ is of semantic type BLOCK.
Based on this distinction, the edge with the correct label is declared in the scope graph.
Using these two inference rules, an inference rule for blocks similar to the one for enumerations is defined.

\begin{align*}
  \frac
  {\nabla s_b \quad s_b \xrightarrow{P} s_r \quad T \equiv \text{BLOCK}(i,n,s_b) \quad s_r \dat{\colon} (i,T) \quad \forall_{Property(i_p,n_p,t_p) \in P}\ {Property(i_p,n_p,t_p)_\ok}}
  {\text{Block}(i,n,P)_\ok}
\end{align*}

\subsection{Transforming in Stratego}
The transformation of the XMI AST to a mCRL2 AST is performed using term rewriting in Stratego.
In Stratego, typed term rewriting strategies can be formulated.
Everything is represented internally by terms, e.g. the syntactic constructors in AST nodes, AST annotations, scope graph nodes, edges, declarations, and semantic types.
Usually, the source AST is traversed and transformed to the target AST.
As the structure of the XMI AST is completely different than the target AST, some deviations are made.
An example of such deviation follows.

Sort definitions are not created by traversing the AST for the definitions.
Instead, they are created by querying the scope graph and translating the semantic types instead of the syntactical constructors.
Consider the strategy \texttt{block-to-mcrl2} in Figure \ref{fig:stratego-block}.
This strategy transforms a term of the type of semantic types TYPE to a term of syntactic type MCRL2-SortDecl.
The latter type is a syntactical constructor part of the mCRL2 syntax specification.

\begin{figure}[!ht]
  \centering
\begin{lstlisting}
block-to-mcrl2(|stx, Hashtable, IndexedSet) :: TYPE -> MCRL2-SortDecl
block-to-mcrl2(|stx, table, set) :
    BLOCK(i, n, s) -> SortAlias(n', Struct([ConstrDecl(n', [l'], [])]))
  with
    n' := <id-to-mcrl2; unique-var-name(|stx, table, set)> n;
    l := <query-var(|stx)> s;
    l' := MCRL2-ConstrDecl-Projs(<map(block-var-to-mcrl2(|stx, table, set))> l)
  \end{lstlisting}
  \caption{Term rewriting strategy for semantic type BLOCK.}
  \label{fig:stratego-block}
\end{figure}

The strategy is specified in Stratego term rewriting rules by applying other term transformation strategies.
A unique but consistent name $n'$ is generated, \texttt{query-var} queries the scope graph for the data declarations in scope $s$ which are also transformed to mCRL2 using the \texttt{block-var-to-mcrl2} strategy.
An interesting rule is the declaration of term \texttt{l}.
The \texttt{query-var} strategy, which is passed the static analysis results in term \texttt{stx}, queries the scope var for the datums of relation $\colon$ in scope $s$.
Applying this strategy to the blocks in Figure \ref{fig:itemtypes-bdd}, the mCRL2 sort specification in \ref{fig:itemtypes-mcrl2} is produced.

\begin{figure}[!ht]
  \centering
  \begin{lstlisting}
sort
  SetOpAutobediening = struct SetOpAutobediening(bepaaldheid4 :eBepaaldheid);  
  SetOpHandbediening = struct SetOpHandbediening(bepaaldheid5 :eBepaaldheid);
  \end{lstlisting}
  \caption{The generated mCRL2 sort specification corresponding to blocks Figure \ref{fig:itemtypes}.}
  \label{fig:itemtypes-mcrl2}
\end{figure}

% Needed as the figure is placed in the Dezyne appendix due to page overflow.
\newpage
\ 

\newpage
\section{Dezyne pull-style models}
\label{app:dzn-pull}
In the Dezyne pull-style models, components only retain their own local state.
They do not store any information which can be retrieved from other components.
In order to perform their desired behaviour, the components must poll other components their state before being able to execute the right commands.

The interface definitions of the controlling component and the feedback loops must be merged, as Dezyne does not permit circular port bindings.
As the feedback is now only supplied on demand, both the actions in the original controlling interface and the status updates in the feedback loop interface are defined as in-events on the interface provided by the controllable component.
The controlling component requires this interface, and thus can query the status and executes the commands on the controllable component.

Consider again the interface definitions containing the \texttt{SetOpHandbediening} command and the \texttt{SetOpAutobediening} command depicted in Figure \ref{fig:itemtypes-bdd}.
Assume that the component is also able to query the current control mechanism (\texttt{bedieningswijze} in Dutch) of the controlled component.
The Dezyne representation of such interface is given in Figure \ref{fig:dzn-pull-if}.

\begin{figure}[!ht]
  \centering
  \begin{lstlisting}
import enums.dzn;

interface bed_BF_Overdruk_Veilige_Ruimte
{
  // control
  in void setOpHandbediening();
  in void setOpAutobediening();

  // feedback
  in eBedieningswijze bedieningswijze();

  behavior {
    // control
    on setOpHandbediening: ;
    on setOpAutobediening: ;

    // feedback
    on bedieningswijze: reply(eBedieningswijze.AUTO);
    on bedieningswijze: reply(eBedieningswijze.HAND);
  }
}
  \end{lstlisting}
  \caption{Example Dezyne interface snippet for the pull-style models.}
  \label{fig:dzn-pull-if}
\end{figure}

According to this interface specification, the controlling component cannot directly observe any change in state as a result of the commands.
The reason for the choice of this representation is that the component must not push its state to this component.
Consider, for example, a second component controlling the control mechanism of this component.
Even if we just sent one of the control mechanism commands, some other component might already have overwritten it.
Therefore, on any further interaction, we must poll the control mechanism using the \texttt{bedieningwijze} event.
An example component snippet in this modelling style which controls the interface defined in Figure \ref{fig:dzn-pull-if} is given in Figure \ref{fig:dzn-pull-comp}.

\begin{figure}[!ht]
  \centering
  \begin{lstlisting}
import enums.dzn;
import bed_BF_Overdruk_Veilige_Ruimte.dzn;
import some_interface.dzn;

component controller
{
  provides some_interface some_if;
  requires bed_BF_Overdruk_Veilige_Ruimte bed_BF;

  behavior {
    on some_if.switchToManual(): {
      eBedieningswijze current = bed_BF.bedieningwijze();
      if (!current.HAND) {
        bed_BF.setOpHandbediening();
      }
    }
    on some_if.switchToAuto(): {
      eBedieningswijze current = bed_BF.bedieningwijze();
      if (!current.AUTO) {
        bed_BF.setOpAutobediening();
      }
    }
  }
}
  \end{lstlisting}
  \caption{Example Dezyne component snippet for the pull-style models.}
  \label{fig:dzn-pull-comp}
\end{figure}

In multiple cases in the model, polled values are needed in several stages of the computation.
In all such cases, the computation functions in Dezyne are parameterised, and the initial in-event which starts the thread of execution in the Dezyne component is made responsible for polling all the required state for dealing with said event.
The polled values are passed as parameter to the functions to guarantee that, within the thread of execution of handling the event, all polled values are consistent and the amount of message passing between components is minimised.

The verification performed by Dezyne of proper implementations in components is rather trivial for such interface definitions.
The interface specifications are trivially deadlock free as it is always possible to send commands and status queries.
Given that the interface behaviour specification is stateless, Dezyne merely has to check whether the events defined in the interface are legal to receive at all times, and all replies to said events are guaranteed to be sent with a value defined in the interface.

\newpage
\section{Dezyne push-style models}
\label{app:dzn-push}
In the Dezyne push-style models, components are aware of the state of the components they directly communicate with over a port.
In order to perform their desired behaviour, the components no longer need to poll other components as they have a decently recent representation of the state of other components.
As the state is not guaranteed to be the truth due to the propagation delay between components, components must be resilient against `unexpected' commands.
As most commands are variable updates, with notifications about said updates, an idempotent implementation suffices in most cases.

Again, consider the interface description given in Figure \ref{fig:dzn-pull-if}, but now for push-style models.
As discussed in the paper, Dezyne does currently not permit passing enumeration data as parameter in the events.
Therefore, the events are duplicated with the data encoded in the event names.
Dezyne uses the \texttt{optional} event to indicate that something might or might not spontaneously happen over an interface.
The representation in the interface is given in Figure \ref{fig:dzn-push-if}.

\begin{figure}[!ht]
  \centering
  \begin{lstlisting}
import enums.dzn;

interface bed_BF_Overdruk_Veilige_Ruimte
{
  // control
  in void setOpHandbediening();
  in void setOpAutobediening();

  // feedback
  out void bedieningswijze_HAND();
  out void bedieningswijze_AUTO();

  behavior {
    eBedieningswijze bedieningswijze = eBedieningswijze.AUTO;

    // control
    [!bedieningswijze.HAND] on setOpHandbediening: { bedieningwijze = eBedieningswijze.HAND; }
    [!bedieningswijze.AUTO] on setOpAutobediening: { bedieningswijze = eBedieningswijze.AUTO; }

    // feedback
    on optional: { bedieningswijze_HAND; bedieningwijze = eBedieningswijze.HAND; }
    on optional: { bedieningswijze_AUTO; bedieningswijze = eBedieningswijze.AUTO; }
  }
}
  \end{lstlisting}
  \caption{Example Dezyne interface snippet for the push-style models.}
  \label{fig:dzn-push-if}
\end{figure}

To remove the need to introduce shadow variables to keep track of the state of the interface, the Dezyne 2.18 release is used for the verification.
In the 2.18 version, Dezyne introducees a feature which allows modellers to reference interface variables in the specification of components.
Using this feature, the component of Figure \ref{fig:dzn-pull-comp} can be specified in the push-style as shown in Figure \ref{fig:dzn-push-comp}.

\begin{figure}[!ht]
  \centering
  \begin{lstlisting}
import enums.dzn;
import bed_BF_Overdruk_Veilige_Ruimte.dzn;
import some_interface.dzn;

component controller
{
  provides some_interface some_if;
  requires bed_BF_Overdruk_Veilige_Ruimte bed_BF;

  behavior {
    on some_if.switchToManual(): {
      if (!bed_BF.HAND) {
        bed_BF.setOpHandbediening();
      }
    }
    on some_if.switchToAuto(): {
      if (!bed_BF.AUTO) {
        bed_BF.setOpAutobediening();
      }
    }
    on bed_BF.bedieningswijze_HAND(): {}
    on bed_BF.bedieningswijze_AUTO(): {}
  }
}
  \end{lstlisting}
  \caption{Example Dezyne component snippet for the push-style models.}
  \label{fig:dzn-push-comp}
\end{figure}

In this implementation, the component accepts incoming events for remote changes of the control mechanism without explicitly handling such change.
By the definition of the interface in Figure \ref{fig:dzn-push-if}, the interface variable changes due to the occurrence of this event.
In the push-style models, it is sometimes necessary to notify other interfaces of this change, such as emitting an out-event on the \texttt{some_if} port.
More specifically, it is sometimes necessary to notify multiple ports of the change of some component state variable due to the change of a required port.
Dezyne forbids emitting man out-event to more than one provides port when dealing with a received out-event of a requires port.
These kind of Y-forks potentially leads to behaviour which is beyond the scope of single component verification, and would thus invalidate the guarantees of Dezyne.
Similarly, Dezyne forbids V-forks caused by emitting an out-event during the handling of an in-event on other port than the one over which the in-event was received.

To solve this problem, the \texttt{defer} statement of Dezyne is used.
This enqueues the execution of some block of statements until at least after the handling of the current event is finished.
Consider the \texttt{bedieningswijze-HAND} event handler in Figure \ref{fig:dzn-push-comp}.
If we were to notify ports \texttt{p1} and \texttt{p2} of this change, this would be modelled in the used modelling style as shown in Figure \ref{fig:dzn-defer}

\begin{figure}[!ht]
  \centering
  \begin{lstlisting}
on bed_BF.bedieningswijze_HAND(): {
  defer() { p1.bedieningswijze_HAND(); }
  defer() { p2.bedieningswijze_HAND(); }
}
  \end{lstlisting}
  \caption{Example use of deferred pushing of state.}
  \label{fig:dzn-defer}
\end{figure}

Now, every deferred execution is interacting with a single port and meets the requirements set by Dezyne.
However, there arises a different problem with such implementation.
As interfaces can now arbitrarily cause the emission of deferred events in a cycle of state change commands, the internal buffers in Dezyne can overflow and verification will fail.
As a remedy to this phenomenon, additional state is introduced in every interface to indicate that the interface is `idle'.
An interface is considered `idle' in this context if the component implementing the interface is finished with executing the deferred statements directly caused by the handler of an in-event on that interface.
The overall structure of behaviour specifications in interface definitions becomes as shown in Figure \ref{fig:dzn-if-idle}, together with the component implementation structure in Figure \ref{fig:dzn-if-idle-comp}.

\begin{figure}[!ht]
  \centering
  \begin{lstlisting}
behavior {
  idle = true;
  // other variables

  [idle] {
    // control commands
    on some_command: { ...; idle = false; }
  }
  [!idle] on inevitable: { done; idle = true; }

  // feedback optional events
}
  \end{lstlisting}
  \caption{Example Dezyne interface snippet for the push-style models with idle state.}
  \label{fig:dzn-if-idle}
\end{figure}

\begin{figure}[!ht]
  \centering
  \begin{lstlisting}
    on bed_BF.bedieningswijze_HAND(): {
      defer() { p1.bedieningswijze_HAND(); }
      defer() { p2.bedieningswijze_HAND(); }
      defer() { bed_BF.done(); }
    }
  \end{lstlisting}
  \caption{Example Dezyne component implementation snippet for the push-style models with done statement.}
  \label{fig:dzn-if-idle-comp}
\end{figure}

While this structure forms a good basis for simulating the models using the Dezyne tooling, verification is still troublesome.
The large combinatorial complexity of all variables in the interface definitions, together with the way the Dezyne semantics are encoded in mCRL2, still yields no results after a week of computations.
This is deemed as infeasible in practice.
The final remedy introduced in the paper, is further decomposing the components.
Similar to the SysML extension, Dezyne components are introduced that are responsible for the computation of individual values.
Then, the original Dezyne component is replaced with a Dezyne system which binds external interfaces to the internal component interfaces.

Using this structure, the verification by the Dezyne tooling has successfully been performed with an event queue and defer queue size of 8.
Still, the tooling is not yet able to explicitly generate the state space of the overall system or verify other than the out-of-the-box provided properties for the overall system.

\end{document}